\documentclass[aps,pra,twocolumn,showpacs]{revtex4}
\usepackage{amsmath}
\usepackage{amssymb,amsfonts,amssymb}
\usepackage[colorlinks=true,dvipdfm]{hyperref}
\usepackage{graphicx}

\begin{document}

\title{Experimental Observation of a Topological Phase in the Maximally
Entangled State of a Pair of Qubits}
\author{Jiangfeng Du$^{1,2}$}
\email{djf@ustc.edu.cn}
\author{Jing Zhu$^{1}$}
\author{Mingjun Shi$^{1}$}
\author{Xinhua Peng$^{2}$}
\author{Dieter Suter$^{2}$}
\affiliation{$^1$ Hefei National Laboratory for Physical Sciences at Microscale and
Department of Modern Physics, University of Science and Technology of China,
Hefei, Anhui 230026, People's Republic of China \\
$^2$Fachbereich Physik, Universit\"{a}t Dortmund, 44221 Dortmund, Germany}
\date{\today}

\begin{abstract}
Quantum mechanical phase factors can be related to dynamical effects or to
the geometrical properties of a trajectory in a given space - either
parameter space or Hilbert space. Here, we experimentally investigate a
quantum mechanical phase factor that reflects the topology of the SO(3)
group: since rotations by $\pi$ around antiparallel axes are identical, this
space is doubly connected. Using pairs of nuclear spins in a maximally
entangled state, we subject one of the spins to a cyclic evolution. If the
corresponding trajectory in SO(3) can be smoothly deformed to a point, the
quantum state at the end of the trajectory is identical to the initial
state. For all other trajectories the quantum state changes sign.
\end{abstract}

\maketitle

Quantum phase factors are ubiquitous and have been crucial in explaining
many phenomena that appear to be unrelated. The overall phase change
resulting from the $2\pi $ rotation of a particle, e.g., distinguishes
Fermions from Bosons or categorizes Anyons \cite{PhysRevLett.49.957}. If
more general circuits are considered than simple $2\pi $ rotations, the
states can acquire arbitrary phases even for Fermions and Bosons \cite%
{3718,453,42}. These phases can be split into two parts, which are referred
to as dynamic and geometric. The dynamic phase is related to the energy
expectation value of the quantum state, integrated over the trajectory,
while the geometric part is related only to the geometry of the circuit.
This analysis of quantum phases in terms of dynamical and geometrical phases
was used extensively over the past decades to discuss a wide range of
quantum phenomena \cite{Bohm:2003aa}.

This picture appears to be not quite complete, however, if we consider the
evolution of pairs of spins 1/2 in a maximally entangled state (MES). If one
of the two spins undergoes a rotation (i.e. a local transformation), the
system acquires neither a dynamical nor a geometrical phase \cite%
{Milman:2006aa}, except at some points, where it abruptly changes by $\pi$
\cite{Milman:2003aa}. It may thus be more appropriate to relate this \emph{%
sudden} phase change to the topology of the appropriate space, i.e. to its
connectedness, rather than to the geometry, i.e. its curvature \cite%
{Milman:2003aa}.

If we consider rotations of single qubits, the appropriate space is that of
the rotations in $R^{3}$, which form a representation of SO(3). Its elements
can be considered to form a sphere, where the direction of each point
corresponds to the rotation axis and the distance from the origin to the
rotation angle. Since rotations by $\pi $ around opposite directions are
indistinguishable in $R^{3}$ and corresponding elements of SO(3) are
identical, opposite points on the surface of the SO(3) sphere have to be
identified. A trajectory that crosses the surface of the sphere immediately
re-enters it at the opposite point. A trajectory that penetrates the surface
once cannot be smoothly deformed into one that does not cross the surface.
Closed trajectories can thus be classified into \textquotedblleft +" and
\textquotedblleft -" classes, depending on the number of times they cross
the surface.

This behavior can be directly mapped into the phases of the quantum states
of maximally entangled spin pairs, where one of the two spins is rotated
around a (possibly time-dependent) magnetic field. For an arbitrary \emph{%
cyclic} sequence $C$ of rotations of the single qubit, i.e. arbitrary
trajectories in SO(3), the MES is transformed into
\begin{equation*}
\vert \Psi \rangle_{MES}(C)\,=\,(-1)^n\,\vert \Psi \rangle_{MES}(0)\,
\end{equation*}
where $\vert \Psi \rangle_{MES}(0)$ is the initial state and $n$ is the
number of times the trajectory penetrates the surface of the SO(3) sphere.
Milman and Mosseri therefore call this phase a topological phase, since it
appears to be related to the double-connectedness of SO(3) \cite%
{Milman:2003aa}.

While Milman and Mosseri considered trajectories consisting of discrete
rotations around fixed axes, LiMing et al \cite{liming:064301} found the
same behavior for trajectories where the rotation axis changes continuously.
They also discussed the possibility of observing the phase in an optical
interference experiment.

In this paper, we report an experimental verification of this topological
phase by NMR interferometry. For this purpose, we initialize a system of two
nuclear spins into a (pseudo-)maximally entangled state and apply
radio-frequency pulses that rotate one of the two spins through trajectories
that correspond either to the ``+" or ``-" type. In the first case, the
resulting signal is identical to that of the reference system, which is not
rotated, in the second case, we observe a phase change by $\pi$.

The topology of SO(3) can, in principle, be explored by letting single
qubits undergo the corresponding rotations. However, in this case, the
resulting phase factors contain dynamical as well as geometrical
contributions \cite{4685,Du:2003aa}. In the present context, Maximally Entangled
States (MES) of two qubits offer a useful alternative. If we initially
prepare the system in a MES and apply local transformations (i.e. rotations)
to one of the two qubits, the system always remains in an MES and does not
acquire any dynamical phase.

A general two-qubit MES can be written as
\begin{equation}
|\Psi \rangle =\sqrt{\frac{1}{2}}(\alpha |00\rangle +\beta |01\rangle -\beta
^{\ast }|10\rangle +\alpha ^{\ast }|11\rangle ),
\end{equation}%
where the coefficients $\alpha $ and $\beta $ are normallized to unity: $%
\alpha \alpha ^{\ast }+\beta \beta^{\ast }=1$. Without loss of generality,
we choose to initialize the system in the Bell state
\begin{equation}
|\Psi\rangle_{MES}(0) =\sqrt{\frac{1}{2}}(|00\rangle +|11\rangle )
\end{equation}
(i.e. $\alpha =1, \beta =0$), and apply the rotations to the first of the
two qubits. At every point in time, we can thus identify the state of the
system
\begin{equation}
|\Psi\rangle_{MES}(t) = \left( U_{\hat{n}}(\theta) \otimes \mathbf{1}
\right) |\Psi\rangle_{MES}(0)
\end{equation}
with the corresponding element of SO(3). Here, $U_{\hat{n}}$ describes the
unitary transformation implementing the trajectory on the first qubit and $%
\mathbf{1}$ is the unit operator acting on the second qubit of the MES. The
unit vector $\hat{n}$ defines the overall rotation axis and $\theta$ the
rotation angle, which corresponds in SO(3) to the distance from the origin.

\begin{figure}[bph]
\centering
\includegraphics[width=0.5\columnwidth]{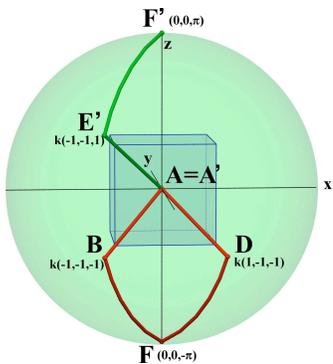}
\caption{(Color online) We consider two different trajectories in SO(3). The
red curve $A\rightarrow B\rightarrow F\rightarrow D\rightarrow A$ belongs to
the \textquotedblleft +" class, since it does not cross the surface of the
sphere. The trajectory $A\rightarrow B\rightarrow F=F^{\prime }\rightarrow
E^{\prime }\rightarrow A^{\prime }$ belongs to the \textquotedblleft -"
class. Only the second part differs from the first trajectory; it is drawn
in green. $k=\frac{2\protect\pi }{3\protect\sqrt{3}}\approx 1.21$. }
\label{class}
\end{figure}

We consider the two types of trajectories shown in Fig. \ref{class}. \cite%
{Milman:2006aa} In each case, the first two rotations take the system from
the origin (point $A$ in Fig. \ref{class}) to point $B$ and from there to
point $F$. The trajectories are chosen such that $F$ corresponds to a
rotation by $\pi$; it is therefore located on the surface of the SO(3)
sphere and is equivalent to point $F^{\prime}$ at the opposite position.
From here, the ``+" trajectory returns to the origin via point $D$, i.e.
without crossing the boundary, but the ``-" trajectory returns via $%
E^{\prime}$. Since it "jumps`` from $F$ to $F^{\prime}$, the associated
quantum state changes sign.

Each of the segments of the trajectories represented in Fig. \ref{class}
correspond to a rotation by $\theta = \frac{2 \pi}{3}$ around one of the
cube diagonals axes. Table \ref{rotate} lists the rotation axes for each
segment of both trajectories.

\begin{table}[th]
\caption{Rotation axes of all segments of the ``+" and ``-" trajectories.}
\label{rotate}%
\begin{tabular}{ll}
\hline\hline
``+" class $ABFDA$ & \multicolumn{1}{|c}{``-" class $ABF^{\prime}E^{%
\prime}A^{\prime }$} \\ \hline
$AB:\sqrt{1/3}\left( -1,-1,-1\right) $ & $AB:\sqrt{1/3}\left(
-1,-1,-1\right) $ \\ \hline
$BF:\sqrt{1/3}\left( 1,-1,-1\right) $ & $BF:\sqrt{1/3}\left( 1,-1,-1\right) $
\\ \hline
$FD:\sqrt{1/3}\left( -1,-1,1\right) $ & $F^{^{\prime }}E^{^{\prime }}:\sqrt{%
1/3}\left( 1,-1,-1\right) $ \\ \hline
$DA:\sqrt{1/3}\left( -1,1,1\right) $ & $E^{^{\prime }}A^{^{\prime }}:\sqrt{%
1/3}\left( 1,1,-1\right) $ \\ \hline\hline
\end{tabular}%
\end{table}

The difference between the two circuits is an overall phase factor acquired
by the quantum state. Since this does not affect directly observable
quantities of the system, one usually resorts to interferometric experiments
for observing the sign change. Milman and Mosseri \cite{Milman:2003aa}
suggested using optical interferometry for this purpose. Here, we resort to
NMR interferometry \cite{4685}.

\begin{figure}[tbp]
\centering
\includegraphics[width=0.95\columnwidth]{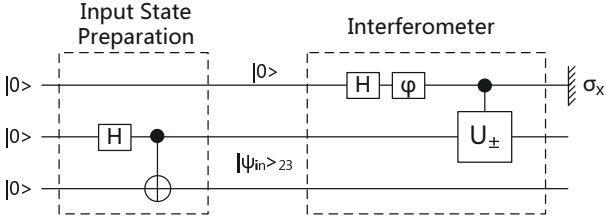}
\caption{Quantum network representation of input state preparation and
interferometric measurement. the initial state is $|000\rangle $. H is a
pseudo-Hadamard gate $H=e^{-i\frac{\protect\pi }{2}I_{y}}$ rotating the
qubit by the angle $\frac{\protect\pi }{2}$ about the Y axis, gate $\protect%
\varphi =e^{-i\protect\varphi I_{z}}$ rotating the qubit by the angle $%
\protect\varphi $ about the Z axis. }
\label{topnet2}
\end{figure}

For this purpose, we have to introduce an ancilla qubit that is coupled to
the two qubits forming the MES. As shown in Fig. \ref{topnet2}, the ancilla
qubit is initialized into an equal superposition by applying a Hadamard gate
to the $\vert 0 \rangle$ state. This Hadamard gate corresponds to the first
beam splitter in a Mach-Zehnder interferometer. After this gate, the system
is in the state
\begin{equation}
|\psi(0) \rangle = \frac{1}{\sqrt{2}}( |0 \rangle + |1 \rangle )
\otimes|\Psi \rangle_{MES} .
\end{equation}
Here, the first qubit is the ancilla qubit and qubits 2 and 3 form the MES.

Instead of the simple unitary operation corresponding to the trajectories in
SO(3), we then use conditional rotations, which only act on that ``copy" of
the MES that is connected to the $\vert 1 \rangle$ state. After this
controlled cyclic circuit, the system reaches the state
\begin{equation}
|\psi(cU_\pm) \rangle = \frac{1}{\sqrt{2}}( |0 \rangle \pm |1 \rangle )
\otimes|\Psi \rangle_{MES} ,
\end{equation}
where the $\pm$ signs refer to the corresponding circuit class. If we trace
over the qubits 2 and 3, we apparently obtain the sign information by
measuring the expectation value of
\begin{equation*}
\langle \sigma_x^1 \rangle = \pm 1 ,
\end{equation*}
where the sign again relates to the circuit class.

As a quantum register for these experiments, we selected the three $^{19}$F
nuclear spins of Iodotrifluoroethylene (F$_2$C=CFI). This system has
relatively strong couplings between the nuclear spins, large chemical
shifts, and long decoherence times. The experiments were performed on a
Bruker Avance II 500 MHz (11.7 Tesla) spectrometer equipped with a QXI probe
with pulsed field gradient. The resonance frequency for the $^{19}$F spins
is around 470.69 MHz. The Hamiltonian of this system is (in angular
frequency units)
\begin{equation}
H=\sum_{i=1}^{3}\omega _{i}I_{z}^{i}+2\pi
\sum_{i<j}^{3}J_{ij}I_{z}^{i}I_{z}^{j},  \label{eq.H}
\end{equation}
where $I_z^i$'s are the local spin operators. The $\omega_i$ are the Larmor
frequencies of the individual qubits. Relevant are the frequency differences
$\omega_1 - \omega_2 \approx 12.02$ kHz and $\omega_2 - \omega_3 \approx
17.33$ kHz, and the coupling constants $J_{12}=64.2$Hz, $J_{13}=51.3$Hz, and
$J_{23}=-129.0$Hz.

\begin{figure}[tbp]
\includegraphics[width=0.99\columnwidth]{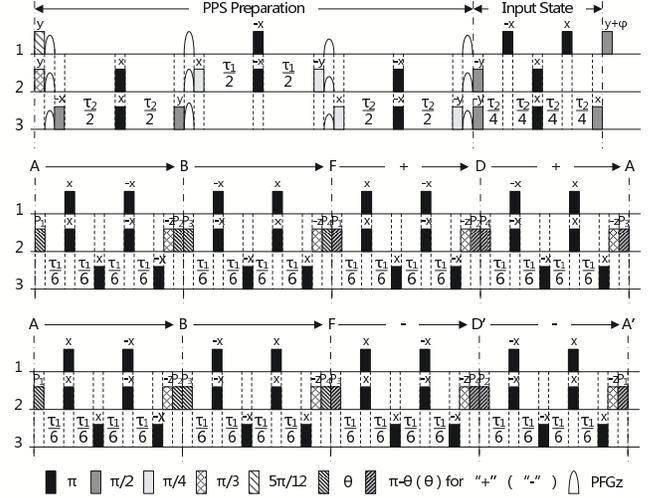}
\caption{Sequence of radio-frequency and field gradient pulses used for the
preparation of the initial state (first line) and to drive the system
through the different trajectories. The second line shows the sequence for
the ``+" trajectory, the third line for the ``-". The parameters are delays $%
\protect\tau_1 = 1/2 J_{12}$ and $\protect\tau_2 = 1/2 J_{23}$, flip angle $%
\protect\theta = \arccos(1/\protect\sqrt{3})$, and phases $P_1= 3 \protect\pi%
/4$, $P_2= 7 \protect\pi/4$, $P_3= 5 \protect\pi/4$, $P_4= \protect\pi/4$. }
\label{f.sequence}
\end{figure}

The system was first prepared in a pseudopure state (PPS) $\rho_{000} =
\epsilon( |000\rangle \langle 000|- \frac{1}{8} \mathbf{1} )$, where $%
\epsilon \approx 10^{-5}$ describes the thermal polarization of the system.
For this initial state preparation, we used spatial averaging \cite%
{Cory:1998aa} by the pulse sequence \cite{Peng:2002aa} shown in the first
line of Fig. \ref{f.sequence}.

From the input state $\vert 000\rangle$, we prepare the maximally entangled
Bell state of qubits 2 and 3 with a Hadamard and a CNOT gate. The ancilla
qubit 1 is then put into a superposition state by another Hadamard gate. The
actual trajectories are implemented by rotating qubit 2, conditioned on the
state of the ancilla qubit. This part of the pulse sequence is represented
in the second and third line of Fig. \ref{f.sequence}. The actual rotation
operations, which are implemented by the gray and hatched pulses, occur only
on the second qubit, while the $\pi$ rotations are applied for refocusing
the coupling to qubit 3 while retaining the coupling with qubit 1.

In order to improve the fidelity of these operations, we implemented the
pulses as robust strongly modulating pulses (SMP) \cite%
{Fortunato:2002aa,Pravia:2003aa,Mahesh:2006aa}. We maximized the gate
fidelity of the individual propagators for a suitable range of radio
frequency field strengths. The theoretical fidelities over the relevant
range of experimental parameters exceeded 0.995 for the individual gates,
and the resulting pulse durations ranged from $200$ to $500$ $\mu s$.

The algorithm requires the measurement of $\langle \sigma _{x}^{1}\rangle $,
the x-component of the ancilla qubit. In an NMR experiment, this corresponds
to the first point of the free induction decay (FID) In practice, better
results are obtained by recording the complete FID, Fourier-transforming it
and integrating the signal over the relevant frequency range.

\begin{figure}[htbp]
\centering
\includegraphics[width=0.9\columnwidth]{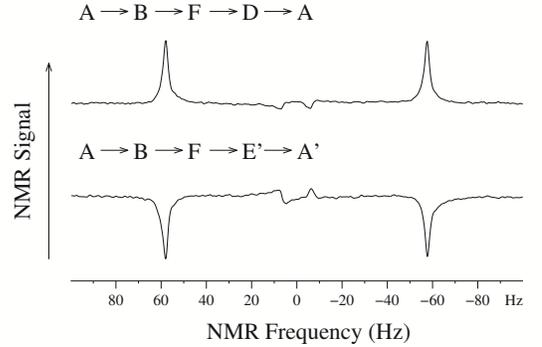}
\caption{$^{19}$F spectra of the ancilla qubit. The upper spectrum was
obtained after the ``+" trajectory was applied to qubit 2, the lower one
after the ``-" trajectory.}
\label{spec}
\end{figure}

In the experimental spectra, shown in Fig. \ref{spec}, the resonance line is
split by the coupling to the second qubit. While the algorithm only requires
the measurement of the integrated signal, the lineshapes and the relative
amplitudes provide useful additional information about the quality of the
measurement. Ideally, both resonance lines should have absorption lineshapes
and the amplitudes should be equal. Obviously, the experimental data agree
well with these predictions.

The upper spectrum was obtained after applying the ``+" trajectory (red
curve in Fig. \ref{class}). In this case, the signal amplitude is positive,
indicating that the trajectory did not change the phase of the quantum
state. In the lower trace, we show the corresponding data after the system
underwent the conditional ``-" trajectory. In this case, the signal is
inverted, as expected for a $\pi$ phase change.

As an additional check that the observed sign change arises from a phase
angle acquired during the trajectory, we measured a complete interferogram,
by shifting the relative phase of the two states of the ancilla qubit and
measuring the signal for each phase value.

\begin{figure}[tbph]
\centering
\includegraphics[width=1\columnwidth]{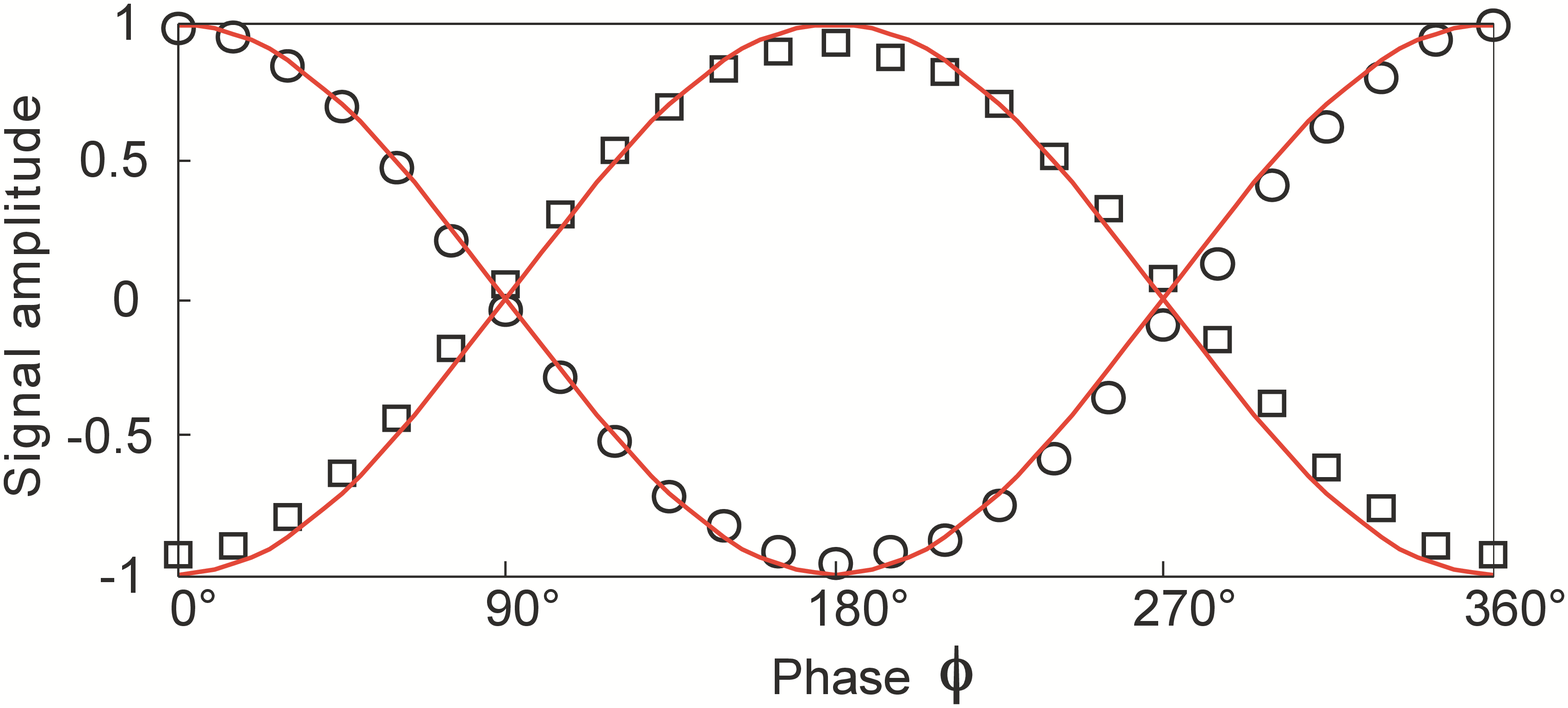}
\caption{(Color online) Experimental interference patterns. The squares
represent the experimental data points for the \textquotedblleft -"
trajectory, the circles those of the \textquotedblleft +" class. The lines
represent the corresponding theoretical functions.}
\label{interferogram}
\end{figure}

Figure \ref{interferogram} compares the experimentally observed signal
amplitudes to the theoretical curve
\begin{equation*}
\langle \sigma_x^1 \rangle = \cos (\varphi - \gamma_{\pm}).
\end{equation*}
$\varphi$ is the experimentall introduces phase shift, which corresponds to
a delay in one arm of a Mach-Zehnder interferometer and $\gamma_{\pm} = \{0,
\pi\}$ is the phase change due to the circuit. The agreement between the
theoretical and experimental data is quite satisfactory and clearly verifies
the expected phase shift of $\pi$ for the ``-" trajectory.

When a quantum state undergoes a cyclic trajectory, it acquires a phase
factor that includes a geometrical part \cite{3718,453,42,Bohm:2003aa}. This
geometrical phase is given by the total curvature of the surface enclosed by
the circuit. A small variation of that circuit leads therefore, in general,
to a small change of the geometrical phase.

The situation is different in the present case: small variations of the
trajectory do not change the overall phase factor \cite{liming:064301}.
Instead, we only have two classes of trajectories: if the trajectory crosses
the surface of the SO(3) sphere an even number of times (including 0), the
total phase vanishes; if the number of crossings is odd, the state reverses
its sign. The different behavior of these two classes of trajectories is
directly related to the double connectedness of SO(3), and the observed
phase factor may therefore be called a topological phase. A related
situation is that of conical intersections \cite{2211,3340}, where the phase
change does not depend on the area enclosed by the circuit, but only by the
number of times it encircles the point of intersection. Possible extensions
of this work include the investigation of multi-qubit systems for different
degrees of entanglement and noncyclic evolutions. These results may be
relevant for topological quantum computation \cite{A.-Yu.-Kitaev:2003aa,
bombin:160502}.

\begin{acknowledgments}
We acknowledge the support by National Natural Science Foundation of China,
the CAS, Ministry of Education of PRC, and the National Fundamental Research
Program. This work is also supported by the DFG under contract Su 192/19-1,
and by the European Commission under contract No. 007065 (Marie Curie
Fellowship).
\end{acknowledgments}

\bigskip

\textit{Note added.} After we finished the experiments, we became aware of a
related optical experiment \cite{Souza:2007aa}.

\bibliography{bibliography}

\end{document}